\documentclass[twocolumn,twocolappendix]{aastex63}
\usepackage[utf8]{inputenc}
\usepackage{placeins}
\usepackage{graphicx}
\usepackage{amsmath}
\usepackage{physics}
\usepackage{mathtools}

\usepackage{dcolumn}
\newcolumntype{d}[1]{D{.}{.}{#1}}
\begin{document}

%\large

\title{Weighing the Milky Way's Satellite Galaxies Using Pulsar Accelerations}

%add orcid ids
\author[0000-0002-7746-8993]{Thomas Donlon II}
\affiliation{Department of Physics and Astronomy, University of Alabama in Huntsville, 301 North Sparkman Drive, Huntsville, AL 35816, USA}
\correspondingauthor{Thomas Donlon II}
\email{thomas.donlon@uah.edu}

\author[0000-0001-6711-8140]{Sukanya Chakrabarti}
\affiliation{Department of Physics and Astronomy, University of Alabama in Huntsville, 301 North Sparkman Drive, Huntsville, AL 35816, USA}

\author[0000-0001-8917-1532]{Jason A. S. Hunt}
\affiliation{School of Mathematics \& Physics, University of Surrey, Stag Hill, Guildford, GU2 7XH, UK}

\begin{abstract}
The properties of dwarf galaxies orbiting the Milky Way (MW) are useful for testing models of the formation of our Galaxy, and by extension various theories of cosmology. Recent efforts to measure the masses of the MW's satellite dwarf galaxies have relied on the motions and positions of stars in the MW's disk and halo, which are perturbed by the passage of satellite galaxies. As there are many known processes in our Galaxy that lead to observed disequilibrium in stars, these kinematic methods have been limited by the inherent difficulty in identifying only the perturbations due to particular satellite galaxies. To bypass some of these restrictions, we present a novel method for determining the masses of two MW satellite galaxies -- the Large Magellanic Cloud (LMC) and the Sagittarius Dwarf Spheroidal Galaxy (Sgr dSph) -- using only direct, instantaneous acceleration data derived from extremely precise timing of millisecond pulsars near the Sun. As the LMC and Sgr dSph orbit the MW, they cause wave-like distortions in the structure of the disk plus a large-scale offset in the centers of mass of the dark matter halo and the baryonic disk. These two effects lead to asymmetric accelerations above and below the disk midplane near the Sun, which is observed in the pulsar acceleration data. Notably, the amplitude of this asymmetry is shown to depend on the masses of the orbiting satellites. We analyze a grid of simulations with varying masses of each satellite in order to determine how well the observed acceleration asymmetry is reproduced by each pair of satellite masses. We find the total (dark + baryon) mass enclosed within the tidal radius at the present day for the LMC to be 4.1 $\pm$ 1.0 $\times$ 10$^{10}$ M$_\odot$ within a radius of 16.6 kpc, and for Sgr to be 3.5 $\pm$ 2.4 $\times$ 10$^8$ M$_\odot$ within a radius of 5 kpc. These results are generally consistent and competitive with previous simulations and determinations of the masses of these objects, but entirely independent of any stellar kinematic data for the first time. \\\vspace{0.5cm}\end{abstract}

\section{Introduction}

%TODO: Add explicit table of pulsar accelerations?

Galaxies grow over time through mergers with smaller dwarf galaxies \citep{DekelSilk1986,Frenk1988,Somerville2015}, a process which is still ongoing in the Milky Way (MW). As a result, the distribution and properties of the MW's satellites at the present day are useful tools for testing ideas about how our Galaxy formed, such as its dynamical accretion history and environment, as well as providing strong constraints on theories of cosmology \citep{DolivaDolinsky2025}. The MW is currently interacting with several satellite galaxies \citep{Helmi2020}, most notably the Large Magellanic Cloud \citep[LMC,][]{Alves2004,PetersenPenarrubia2021} and the Sagittarius dwarf spheroidal galaxy \citep[Sgr dSph,][]{Ibata1994,Newberg2002,Fardal2019}. 

There is a strong body of evidence that the MW has been disrupted by the passage of its orbiting satellites \citep{HuntVasiliev2025}. These disruptions include waves and ripples induced by a dwarf galaxy passing near the disk \citep{EdelsohnElmegreen1997,Chakrabarti_Blitz2009,Quillen2009,Purcell2011,Laporte2019}, and large-scale structures that arise from bulk tidal motion of the dark matter halo due to the orbit of a massive satellite \citep{Garavito-Camargo2021,PetersenPenarrubia2021,Chandra2025}. The orbit of the Sgr dSph passes through the disk, which is likely linked to the observed phase-space spiral (\citealt{Antoja2018,Laporte2019,Hunt2021}, but see \citealt{BennettBovy2021,Bennett2022}) as well as numerous identifications of disk waves \citep{Xu2015,Poggio2025}. Similarly, the gravitational influence of the LMC is believed to be related to the observed large-scale reflex motions of the stellar halo \citep{PetersenPenarrubia2021}, disruptions to tidal streams \citep{Erkal2019,Vasiliev2021}, and the warp of the stellar disk \citep{Garcia-Ruiz2002}. Each of these processes depends on the properties of the MW as well as the orbit and properties of its orbiting satellites. By studying these various dynamical features, one can gain insight into the structure and history of the MW and its satellite galaxies. 

The structure of the MW has historically been determined by observing the positions and motions of stars within our Galaxy \citep{BlandHawthornGerhard2016} and its satellites \citep{Callingham2019}. A prominent example of this is Jeans modeling, which has been used for nearly a century to determine the surface density of the MW disk (which provides an estimate of the local density of dark matter) from the vertical velocity dispersion of nearby stars \citep{Oort1932,KuijkenGilmore1991,McKee2015}. Similarly, Jeans modeling has been used to constrain the masses of dwarf galaxies \citep[e.g.][]{Strigari2008}. As we cannot directly measure the gravitational forces on stars in the MW or nearby galaxies, these techniques must instead combine kinematic information of stars with assumptions about the state of the Galaxy in order to estimate accelerations, which can then be used to constrain the Galaxy's fundamental properties. These assumptions, such as time-independence and azimuthal or spherical symmetry, are no longer believed to be realistic given the sizable body of evidence for prominent disequilibrium processes throughout the MW, and can introduce serious systematic uncertainties in kinematic studies \citep{Banik2017,Haines2019,Cheng2024}. 

To understand why this is the case, consider the orbit of a single star. The present-day position and velocity of that star are the sum of all gravitational forces it has felt from many dynamical effects over its entire lifetime; this makes it difficult to isolate the signal of a single particular dynamical process when using kinematic data. The star's present-day velocity is an integration over time of the force on that star due to disk waves, plus the phase-space spiral, plus resonances with the bar, plus dark matter substructure, and so on. This ultimately limits both the temporal and spatial resolution with which kinematic studies are able to constrain the MW gravitational field. As such, kinematic data is unable to determine the present-day accelerations across the disk to high accuracy, and must rely on techniques which estimate these accelerations instead.

One solution to this problem is to directly measure the present-day acceleration at different points in the Galaxy, and then use these direct acceleration measurements to infer the current dynamics of the MW. Such an approach avoids any assumptions about time-independence or symmetries of the underlying distribution function -- a key advantage of direct acceleration measurements over kinematic studies \citep{Chakrabarti2021}. This has recently become possible due to the substantial production of pulsar timing array data \citep[e.g.][]{IPTADR2}; by precisely timing the arrivals of a pulsar's radio pulses, it is possible to determine that object's line-of-sight acceleration \citep{Chakrabarti2021,Moran2024,Donlon2024,Donlon2025,Chakrabarti2025}. 

This method works because each disequilibrium effect is expected to lead to instantaneous accelerations in disk stars. For example, a dwarf galaxy passing near the MW disk will lead to wave structure in the disk stars and gas; these local variations in density propagate through the disk stars' self gravity, and material is pulled towards regions of greater density. Satellite galaxies can also lead to large-scale departures from a smooth underlying dark matter halo, either due to tidal stripping distributing the dwarf galaxy's material inhomogeneously throughout the halo, a dark matter wake, or reflex motion of halo stars \citep{Darragh-Ford2025}. These features lead to bulk accelerations towards more dense regions of the halo and away from less dense regions. As the strength of the dynamical perturbations depends on the mass of the satellite galaxy, the associated accelerations will be larger for more massive satellites, all other things held constant. Therefore, if one knows the acceleration field within the MW disk, it is possible in principle to use this information to constrain the masses of the various satellites in the MW.

\begin{figure*}
    \centering
    \includegraphics[width=\linewidth]{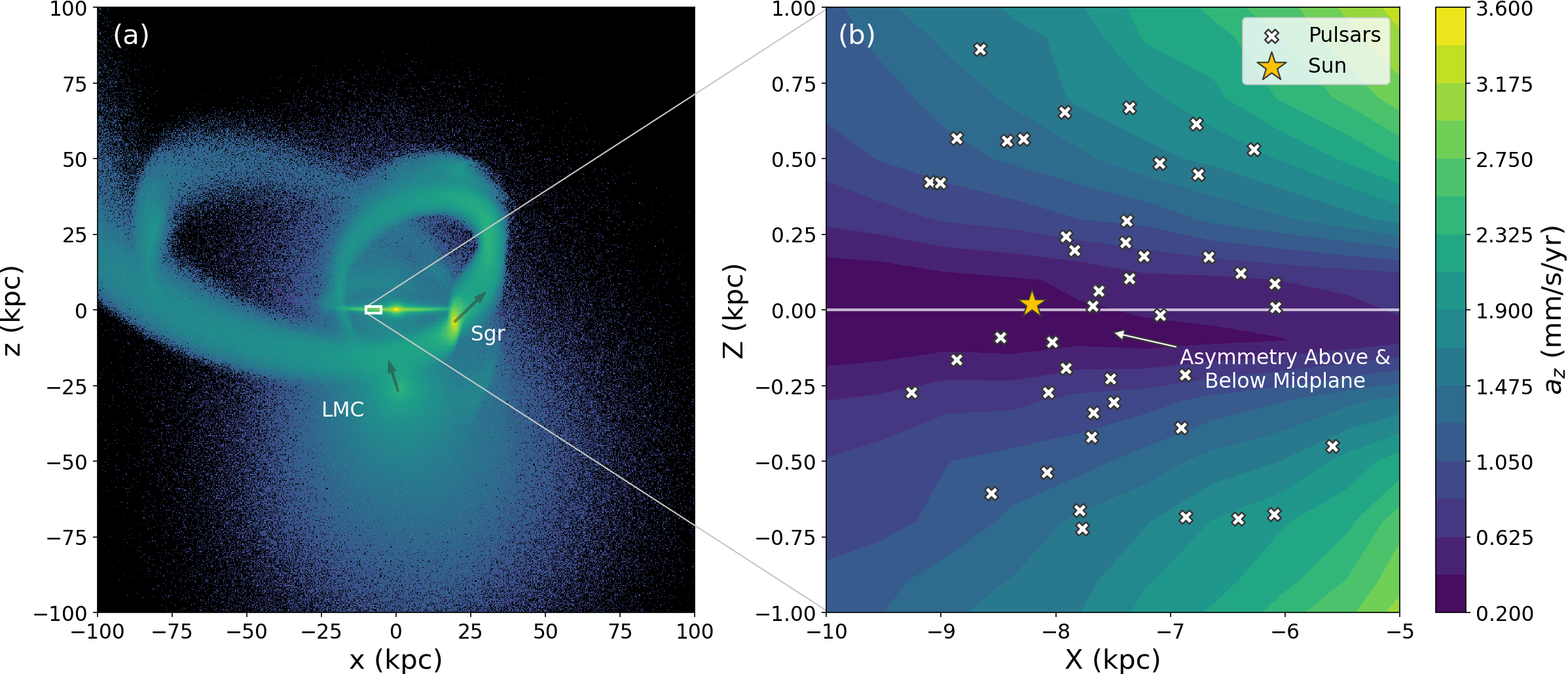}
    \caption{Panel (a): Simulation of the Milky Way, the LMC and the Sgr dwarf galaxy/tidal stream. The density of the dwarf galaxies and their stripped material have been enhanced compared to the host galaxy to make them more visible. The locations and velocities (shown as gray arrows, times a normalization factor) of the orbiting satellites match their observed present-day locations. Panel (b): The locations of the pulsar data compared to the Sun, plotted on top of the simulated vertical accelerations (shown as colored contours). The accelerations above the disk are stronger than the accelerations below the disk; near the midplane, this is shown by the ``lopsided'' contours on either side of the white horizontal line.}
    \label{fig:fancy_pic}
\end{figure*}

With this idea in mind, we aim to use the measured pulsar accelerations to constrain the properties of the MW's satellites. The pulsar accelerations span a relatively small region of the disk (only a few kpc out from the Sun), so it is necessary to restrict our analysis to this region. Simulations of MW-like galaxies have previously suggested that the vertical acceleration profile near the Sun can become asymmetric -- that is, accelerations towards the midplane are stronger on one side of the Galactic disk than the other -- when the galaxy interacted with satellite galaxies \citep{Chakrabarti2020}.  The observed MW pulsar accelerations are indeed found to be vertically asymmetric; they are stronger above the Galactic midplane than below \citep{Donlon2024,Donlon2025}. The size of this effect was initially believed to be as large as 50$-$100\% of the overall acceleration based on measurements of binary millisecond pulsars, although follow-up using an expanded pulsar sample and more realistic modeling procedures found the strength of the asymmetry was closer to 30\% of the overall acceleration. Crucially, \cite{Donlon2025} pointed out that the effects of the LMC and Sgr on the MW could potentially lead to the observed asymmetry because they perturb the structure of the disk and generate distortions in the dark matter halo, leading to large-scale accelerations at levels in that work, it remains to use the observed acceleration data to produce information about the properties of those satellite galaxies. 

Here, we show that the observed asymmetry can in-fact be used to place useful constraints on the masses of the MW's satellite galaxies. As a proof of concept, we focus our analysis to the two satellites that excite the strongest disequilibrium features in the MW: the LMC and Sgr. While other dwarf galaxies may also generate accelerations on the pulsar sample, these accelerations are expected to be small compared to the effects of the LMC and Sgr, as these two galaxies are believed to represent the majority of the MW's present-day features from satellite interactions. While our results are broadly consistent with previous analyses which fit to stellar kinematic and/or dynamical data, here we are able to constrain the masses of the LMC and Sgr dwarfs using only direct acceleration data for the first time. This analysis opens a new avenue for using direct acceleration measurements to measure the properties of the MW and dark matter-dominated dwarf galaxies in the Local Volume.

\section{Pulsar Accelerations}

% We briefly explain here the procedure for determining the acceleration of a millisecond pulsar, and we direct the reader to previous works for further details on this topic \cite{Chakrabarti2021,Donlon2024,Donlon2025}. 

In general, an acceleration can be obtained for an object emitting a periodic signal by considering the time derivative of its Doppler shift, \begin{equation} \label{eq:doppler}
    \frac{\dot{P}}{P} = \frac{a_\mathrm{los}}{c},
\end{equation} where $P$ is the period of the signal, $a_\mathrm{los}$ is the observed line-of-sight acceleration, and $c$ is the speed of light. Note that this acceleration is relative to the Sun, which is also accelerating relative to the Galaxy. As a result, the observed line-of-sight acceleration of an object due to the gravitational field of the MW is actually \begin{equation}
    a_\mathrm{los} = (\mathbf{a} - \mathbf{a}_\odot)\cdot \hat{\mathbf{d}},
\end{equation} where $\mathbf{a}$ is the acceleration of the object in the rest frame of the Galaxy, $\mathbf{a}_\odot$ is the acceleration of the Sun, and $\hat{\mathbf{d}}$ is the unit line-of-sight vector from the Sun to the object of interest. In this work, we use the catalog of observed accelerations for 53 pulsars provided by \cite{Donlon2025}, and provide below a description of how these accelerations are obtained. The positions of these pulsars relative to the Sun and the rest of the MW are shown in the right panel of Figure \ref{fig:fancy_pic}.

Pulsar timing arrays collect times of arrival (TOAs) for the observed radio pulses from pulsars, and use small variations in these TOAs to construct a timing solution for each pulsar \citep[e.g.][]{IPTADR2}. These timing solutions contain all statistically significant relevant information for that pulsar, including the spin period and time derivative of the spin period, proper motion, and parallax of that pulsar. If a pulsar is in a binary system, the timing solution may also contain its binary period, time derivative of the binary period, the orbital eccentricity, and masses of both objects from measurements of the system's Shapiro delay \citep[for example,][]{Cromartie2020}. 

In principle, the pulsar's binary period and its spin period could be used interchangeably in Equation \ref{eq:doppler} to obtain a value for the acceleration of that object. In practice, however, there are several limitations to this process. The observed drifts in the binary and spin periods of the pulsar include contributions from several physical effects, only part of which is due to the underlying gravity of the Galaxy. In the case of the binary orbital period, the emission of relativistic gravitational waves will lead to an decrease in the orbital period over time (i.e. \citealt{WeisbergHuang2016}). In the case of the pulsar's spin period, there is an intrinsic spindown that occurs as the pulsar loses energy and angular momentum, which does not have a suitable description derived from first principles; however, this contribution can be empirically calibrated away for pulsars with low surface magnetic field strengths \citep{Donlon2025}. Both the spin and orbital periods are modulated by the apparent motion of the pulsar on the sky, which is known as the Shklovskii effect \citep{Shklovskii1970}. Once all of the relevant effects are computed from the properties of each pulsar, the true change in the period due to the Galaxy's gravitational field can be obtained. 

In order to provide a constraint on the gravitational field of the Galaxy, we must know the 3-dimensional location for each pulsar along with its line-of-sight acceleration. For this reason, each pulsar in the dataset is required to have a measured parallax. Finally, all pulsars that are interacting in some way with their companions, such as redback, black widow, or other peculiar pulsars \citep[for example,][]{Donlon2025b}, are removed from the dataset, as these objects can have complicated additional contributions to their observed period drifts which are not easily removed from their timing solutions. For further information regarding the details of the pulsar catalog used in this work, we direct the reader to \cite{Donlon2025}, along with \cite{Donlon2024} and \cite{Chakrabarti2021}. The catalog of pulsar accelerations is publicly available at \url{https://github.com/thomasdonlon/Empirical_Model_MSP_Spindown_Accels}. 

Note that for this study we include the intrinsic scatter in the surface magnetic field of the millisecond pulsars as per the appendix of \cite{Donlon2025}. The addition of this noise term is a more realistic treatment of the uncertainty in the accelerations calculated using the spin information of pulsars.

\section{Simulations} \label{sec:simulations}

\subsection{Initial Conditions}

To explore the accelerations induced by satellite galaxies orbiting the MW, we ran a suite of self-consistent $N$-body simulations which included a MW-like host galaxy plus two major satellites representing the LMC and Sgr (left panel of Figure \ref{fig:fancy_pic}). This suite is based on simulations performed by \cite{Stelea2024}, which were constructed to study analogs of the LMC and Sgr around a MW-like host galaxy, and their initial conditions followed that of \cite{Vasiliev2021}. In these simulations, the observed morphology of the Sgr stream and LMC were simultaneously reproduced within the \cite{Vasiliev2021} model in a fully live setting for the first time, while also broadly matching the shape and mass of the MW as assessed via the behavior of the rotation curve. The general shape and mass profile of our MW model is identical to that work, except that we reduced the velocity dispersion of the disk compared to \cite{Stelea2024} so that the disk is similar in scale height to the actual MW disk (without changing its mass). This ensures that we are measuring realistic vertical accelerations throughout the disk stars compared to the real MW, and that these structures interact and dissipate on the correct timescales. The new set of initial conditions for these simulations were generated using the \textit{agama} software \citep{agama} and were integrated for 3 Gyr using the \textit{Gadget-4} software \citep{gadget4}. 

\begin{table}[]
    \centering
    \begin{tabular}{lrrrr} \hline
        \textbf{LMC} & & \\ \hline
        Model & & & $M_\mathrm{Tot}$ & $r_s$ \\
              & & & (10$^{11}$ M$_\odot$) & (kpc) \\ \hline \hline
        Light & & & 1.0 & 8.5  \\
        Medium & & & 1.5 & 10.8 \\
        Medium-Heavy & & & 1.75 & 11.9 \\
        Heavy & & & 2.0 & 12.9 \\
        Heavy-Plus & & & 2.25 & 13.8 \\ \hline
        
        \textbf{Sgr} & & \\ \hline
        Model & $M_*$  & $r_{s,*}$ & $M_\mathrm{Dark}$ & $r_{s,\mathrm{Dark}}$ \\ 
              & (10$^{8}$ M$_\odot$) & (kpc) & (10$^{8}$ M$_\odot$) & (kpc) \\ \hline \hline
        Light & 1.2 & 0.75 & 20.9 & 6.0 \\
        Medium-Light & 1.7 & 0.9 & 31.4 & 7.0 \\
        Medium & 2.3 & 1.0 & 41.9 & 8.0 \\
        Medium-Heavy & 2.9 & 1.1 & 62.8 & 9.0 \\
        Heavy & 4.7 & 1.25 & 83.7 & 10.0 \\ \hline
        
    \end{tabular}
    \caption{Dwarf galaxy model parameters for the simulations used in this work.}
    \label{tab:model_params}
\end{table}

Although it may be possible to constrain various properties of the satellite galaxies using direct acceleration measurements, such as their orbits or their physical extent, we limit our analysis only to the masses of the satellites as a proof of concept. This limits our analysis to just two dimensions, which significantly reduces the amount of computing time required to obtain a result. To explore the dependence of the accelerations in the Solar neighborhood on the masses of the satellites, we ran an initial grid of 9 medium-resolution simulations, each with a different combination of initial masses and scale lengths for the two satellite galaxies. Each simulation had a light, medium, or heavy total mass for the LMC, and similarly had a light, medium, or heavy mass for the Sgr dSph. The LMC model was a single-component dark matter spheroid, while the Sgr dSph was composed of a baryonic component (represented by a King model) embedded in a dark matter component (represented by a spheroid), each with its own scale radius. The masses and scale radii of each satellite model are given in Table \ref{tab:model_params}. In the medium-resolution simulations, the LMC was composed of 500k particles, and Sgr was composed of 360k stellar particles and 150k dark matter particles.

In all simulations, the host galaxy was represented by the same three component model consisting of a bulge, disk and halo. The bulge consisted of a spheroidal distribution function with a total mass of 1.2$\times10^{10}$ M$_\odot$ and a scale radius of 200 pc; the disk had an exponential profile in radius and an isothermal ($\sech^2$) vertical profile, with a total mass of 5$\times10^{10}$ M$_\odot$, a scale length of 3 kpc, and a scale height of 0.4 kpc; and the halo consisted of a double power law profile with a total mass of 7.3$\times10^{11}$ M$_\odot$ within a cutoff radius of 200 kpc, scale length of 7 kpc, an inner slope of $\gamma=1.0$ and an outer slope of $\beta=2.5$. The halo density was truncated at a distance of 200 kpc from the Galactic center. In the medium-resolution simulations, the host galaxy model had 240k bulge particles, 960k disk particles, and 1.8M halo particles. This MW model has a fairly small total mass of approximately 8$\times10^{11}$ M$_\odot$; this is somewhat less than estimates of the MW mass from the orbits of the Magellanic Clouds \citep[e.g.][]{Craig2022}, but it lies within the range of acceptable masses for the Galaxy from a variety of observations \citep{Wang2020,BobylevBaykova2023}.

\subsection{Dynamical Friction}

As dwarf galaxies orbit their host, they lose energy due to dynamical friction; the rate of energy loss increases with the dwarf galaxy's mass \citep{Chandrasekhar1943}. Additionally, the two satellites interact both with each other as well as the live host galaxy, leading to reflex motion that offsets the relative position of each object relative to the overall center of mass \citep{Vasiliev2021}. This makes the orbits of the satellite in each simulation slightly different when the mass of a satellite is changed, meaning that one or both satellites may not necessarily have the correct, observed position and velocity at the end of the simulation. This is problematic, as we wish to only explore the effects of varying the masses of the two satellites on the simulated accelerations, while keeping their final locations and velocities fixed. 

\begin{figure*}
    \centering
    \includegraphics[width=\linewidth]{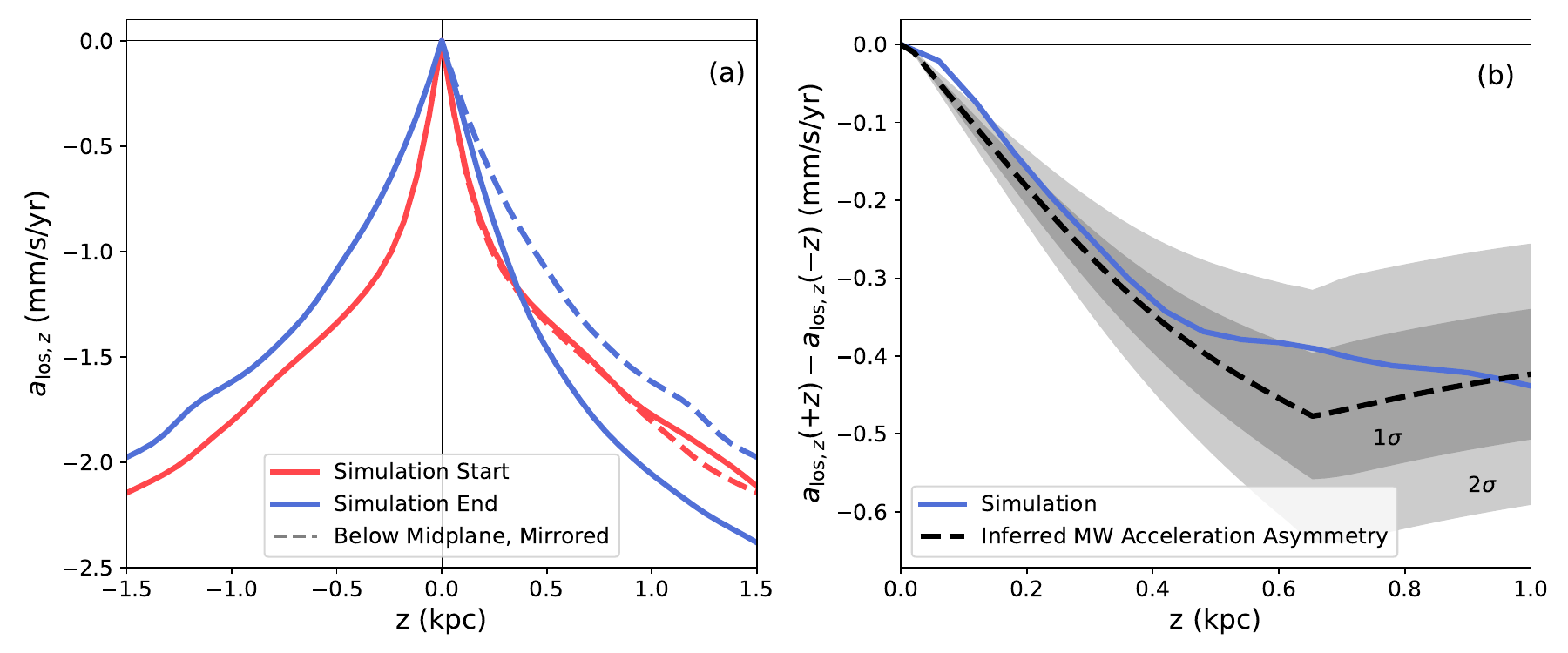}
    \caption{The simulated and observed acceleration asymmetry. Panel (a) shows the vertical component of the line-of-sight acceleration ($a_{\mathrm{los},z}$) for the best-fit simulation above and below the Solar position. The acceleration below the midplane has been mirrored to positive $z$ for easier comparison (dashed lines). This acceleration is shown at the beginning of the simulation (red), where the profile is identical above and below the midplane. At the end of the simulation, after the effects of the satellites have perturbed the galaxy, the acceleration profile is asymmetric, with larger magnitude (more negative) at positive $z$ than at the same distance from the midplane, but negative $z$. This asymmetry can be visualized as the difference between the solid and dashed blue lines, and is plotted in panel (b). The acceleration asymmetry determined from the observed MW pulsar data is also plotted as a black dashed line in panel (b), and the 1$\sigma$ and 2$\sigma$ errors are shown as gray regions around the central value. The simulation agrees with the inferred acceleration asymmetry within this error region. }
    \label{fig:alos_asym}
\end{figure*}

\begin{figure*}
    \centering
    \includegraphics[width=\linewidth]{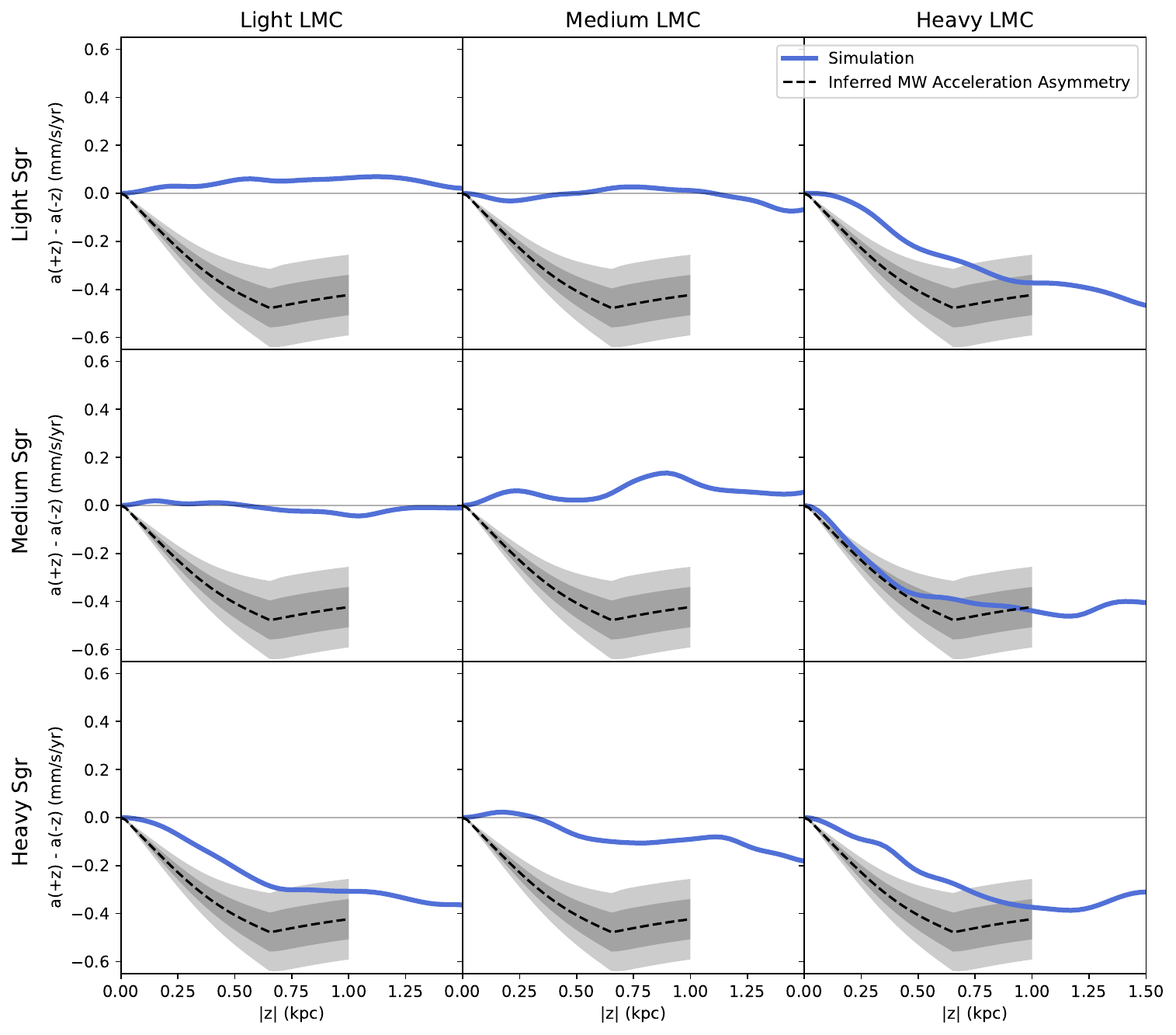}
    \caption{The vertical acceleration asymmetry at the location of the Sun for simulations with different masses of the LMC and Sgr dwarfs. Each column corresponds to a different mass of the LMC, and each row corresponds to a different mass of the Sgr dSph (see Table \ref{tab:sat_masses}). Each panel shows the simulated acceleration asymmetry (acceleration above the disk minus acceleration below the disk) vs. distance from the midplane in blue, and the asymmetry inferred from the observed pulsar data as a dashed black line with 1$\sigma$ error bars. The thin horizontal gray line corresponds to no asymmetry. The simulation that best recovered the MW asymmetry has a heavy LMC and a medium-sized Sgr.}
    \label{fig:supp_asym_grid}
\end{figure*}

Previous works have corrected for this effect by optimizing the orbits of the two satellites using a non-linear method \citep{VasilievBelokurov2020,Stelea2024}, although this is computationally expensive. In our simulations, we compared the fiducial orbits with the orbits in simulations with different satellite masses, and found that the trajectories of the satellite orbits stayed nearly the same in each simulation except for the time it took each satellite to reach its present-day position. More massive satellite models would feel more dynamical friction; this caused them to lose orbital energy more quickly than lighter satellite models, and would therefore have a shorter orbital period. 

In our case, because the effect of dynamical friction was small, we can treat the variations in the travel times of each satellite as a linear correction to the satellites' orbits rather than a nonlinear one (which would require more complicated optimization techniques). We began by integrating the initial condition of each satellite forwards or backwards in time by a small amount as needed (shorter than 100 Myr) in a frozen snapshot potential corresponding to the initial conditions of the host galaxy. The simulation was then integrated forward in time and the final positions of the satellites were determined. This procedure was iterated until the satellites reached their present-day positions simultaneously, which took fewer than five iterations in all cases. Note that the time that the simulated satellites simultaneously reached their present-day positions was not always exactly 3.0 Gyr after the start of the simulation. For example, our fiducial simulation reaches this state only 2.9 Gyr after the simulation begins. As a general rule, we analyze the point in time where the satellites are at their observed present-day positions, regardless of any other factors. This iterative process was performed using low-resolution simulations in order to conserve computing resources, with only 8k bulge particles, 32k disk particles, 60k halo particles, 50k LMC particles, and 36k star and 15k dark particles in the Sgr model. We confirmed that the final locations of the satellites agree between the low and medium resolution simulations -- or in other words, that the dynamical friction in these simulations did not significantly depend on resolution.

\section{Results \& Discussion}

\subsection{Acceleration Asymmetry}

In dynamical equilibrium, the strength of vertical ($z$-component) accelerations on either side of the disk will be equal. Departures from equilibrium can lead to a lopsided density profile, with more stars and/or gas being located on one side of the Galactic midplane \citep{Widrow2012,YannyGardner2013,BennettBovy2019,Guo2022}. As a result, asymmetry in the vertical accelerations on either side of the disk are compelling evidence for disequilibrium, and can potentially be used to constrain the nature of processes that drive lopsided density profiles. 

\cite{Chakrabarti2020} showed using simulations that the vertical acceleration profile will be strongly disrupted by the passage of one or more dwarf galaxies near the Galactic disk. They pointed out that this effect is expected to be large compared to the overall acceleration, and therefore should be easily observable even with a modest amount of direct acceleration data. \cite{Donlon2024} found evidence for this asymmetry in pulsar accelerations for the first time, and showed that it was comparable to the amount of asymmetry predicted by simulations of a MW + Sgr system, although the potential models used in that paper have since been shown to perform poorly on data with a vertical extent larger than a few hundred pc. 

In a follow-up paper, \cite{Donlon2025} produced a clear measurement of the acceleration asymmetry by considering the first few terms of a Taylor series expansion of the observed accelerations (shown in the right panel of Figure \ref{fig:alos_asym}). This was then compared to the observed waves in the stellar disk, plus a MW potential model which included an offset between the centers of mass of the halo and the disk. These two effects, which were chosen because they arise from interactions between the MW and its dwarf galaxies, were able to sufficiently capture the magnitude and shape of the observed acceleration -- strong evidence that the observed acceleration asymmetry is linked to the properties of the LMC and Sgr.  

Returning to our suite of simulations, there is a vertical acceleration asymmetry present which is qualitatively similar to the asymmetry in the observed pulsar accelerations. At the start of the simulation, when the disk is in dynamical equilibrium, the vertical force above and below the disk are identical at all heights, i.e., there is no asymmetry (left panel of Figure \ref{fig:alos_asym}). As the simulation progresses and the effects of the satellites generate disequilibrium in the Galaxy, the acceleration profile becomes lopsided, with stronger forces on one side of the midplane, leading to an asymmetry in the accelerations as observed from the Sun. 

We quantify the strength of the vertical acceleration asymmetry as the difference between the vertical component of the line-of-sight acceleration, $a_{\mathrm{los},z}(z)$, at a distance $z$ above and below the midplane; we then calculate this asymmetry at a range of heights. The direction and magnitude of this asymmetry change over time throughout the simulation due to the differential rotation of the disk stars with respect to the orbiting satellites and the dynamical evolution of waves in the disk. We calculated the difference between the asymmetry inferred from the observed pulsar data and the asymmetry at the Solar position at the end of each simulation; the results are shown in Figure \ref{fig:supp_asym_grid}. The best-fitting simulation is located in the right column and middle row of this figure, corresponding to the simulation where the total mass of each satellites at the beginning of the simulation (3 Gyr ago) is 2.0 $\times$ 10$^{11}$ M$_\odot$ for the LMC, and 4.4 $\times$ 10$^{9}$ M$_\odot$ for the Sgr dSph. The simulated asymmetry has a magnitude of roughly 25\% of the overall vertical force from the disk and practically lies within the 1$\sigma$ standard error bars of the pulsar data, indicating that the observed asymmetry is consistent with the satellite masses in our best-fit simulation. 

% We determine the total mass of each satellite at the beginning of the simulation (3 Gyr ago) to be 2.0 $\pm$ 0.5 $\times$ 10$^{11}$ M$_\odot$ for the LMC, and 4.4 $\pm$ 3.1 $\times$ 10$^{9}$ M$_\odot$ for the Sgr dSph -- our uncertainty in the LMC's total mass is competitive with the precision obtained using kinematic methods. These measurements are based on an observed acceleration asymmetry above versus below the MW disk \cite{Donlon2025}, which arises from distortions in the MW's dark matter halo caused by interactions with the LMC and Sgr. Since the strength and orientation of the acceleration asymmetry depends on the masses of the satellites, we are able to fit the satellite masses using a suite of fully live N-body simulations with different models for the satellites. The best-fitting simulation also reproduces dynamical features such as the phase space spiral and the disk warp. 

% The initial total masses of the satellites in the best-fitting simulations are 2.0 $\pm$ 0.5 $\times$10$^{11}$ M$_\odot$ for the LMC, and 4.4 $\pm$ 3.1 $\times$10$^{9}$ M$_\odot$ for Sgr (details on how these uncertainties are calculated is provided in Methods). 

\subsection{Satellite Mass Uncertainties}

The uncertainties on the best-fit masses were calculated by running a second grid of medium-resolution simulations. This time, the grid was centered on the best-fitting simulation (medium-mass Sgr, heavy LMC) and the mass and scale radius step sizes were cut in half. The goodness-of-fit of each simulation to the observed pulsar data was calculated as \begin{equation}
    \mathcal{L} = \frac{1}{\int_0^z \dd z'} \left( \frac{1}{2}\int_0^z \frac{\left[ \Delta a_\mathrm{sim}(z') - \Delta a_\mathrm{obs}(z')\right]^2}{\sigma^2_{\Delta a_\mathrm{obs}}(z')} \dd z' \right),
\end{equation} where $\Delta a(z) = a_\mathrm{los,z}(+z) - a_\mathrm{los,z}(-z) $ is the vertical acceleration asymmetry at a height of $\pm z$ from the midplane. This integral is essentially equivalent to a $\chi^2$ statistic. The upper bound of the integral in our case was $z=1$ kpc, and consisted of $n=50$ steps in $z$. 

The covariance matrix in each parameter was then estimated as the Cramer-Rao bound, which is obtained by inverting the Hessian of the goodness-of-fit; \begin{equation}
    \mathrm{cov}(\theta_{ij}) = \frac{1}{n}\left[ \frac{\partial^2\mathcal{L}}{\partial \theta_i \; \partial \theta_j} \right]^{-1},
\end{equation} where the partial second derivatives were estimated using finite differences. The reported uncertainties are then \begin{equation}
    \sigma(\theta_i) = \sqrt{\mathrm{diag}(\mathrm{cov}(\theta_{ij}))}.
\end{equation}

From this procedure we determine the total mass of each satellite at the beginning of the simulation (3 Gyr ago) to be 2.0 $\pm$ 0.5 $\times$ 10$^{11}$ M$_\odot$ for the LMC, and 4.4 $\pm$ 3.1 $\times$ 10$^{9}$ M$_\odot$ for the Sgr dSph -- our uncertainty in the LMC's total mass is competitive with the precision obtained using kinematic methods. The relative uncertainty in the mass of Sgr is worse, which is discussed in further detail in Section \ref{sec:sgr_mass}. 

As each simulation is computationally expensive, it was not possible to directly optimize the simulated acceleration asymmetry in order to obtain a better fit to the observed data; it is likely possible to produce a better fit by varying additional parameters of the simulations, such as the model for the host galaxy. Additionally, because of the coarse simulation grid, these reported uncertainties should be treated as estimates. It is also entirely possible that our best-fit simulation is a local maximum of the likelihood surface, and that a better global maximum fit exists either in-between the steps of our grid, or outside of our grid's bounds. These issues could be resolved in the future with a finer grid, and/or a full Bayesian treatment of the problem. Regardless of these caveats, our results here are a demonstration that it is in-fact feasible to constrain the masses of these satellites at a productive level using only direct acceleration measurements as inputs. 

%Our estimate for the mass of Sgr has a larger relative uncertainty, which we speculate may be due to two reasons: first, the lower present-day mass of Sgr compared to the LMC means that it will induce smaller accelerations than the LMC, although this is cancelled out at some level by the fact that the Sgr dSph is closer to the MW disk, and (ii) the effect of the Sgr 

\subsection{Literature Comparison}

\begin{figure*}
    \centering
    \includegraphics[width=\linewidth]{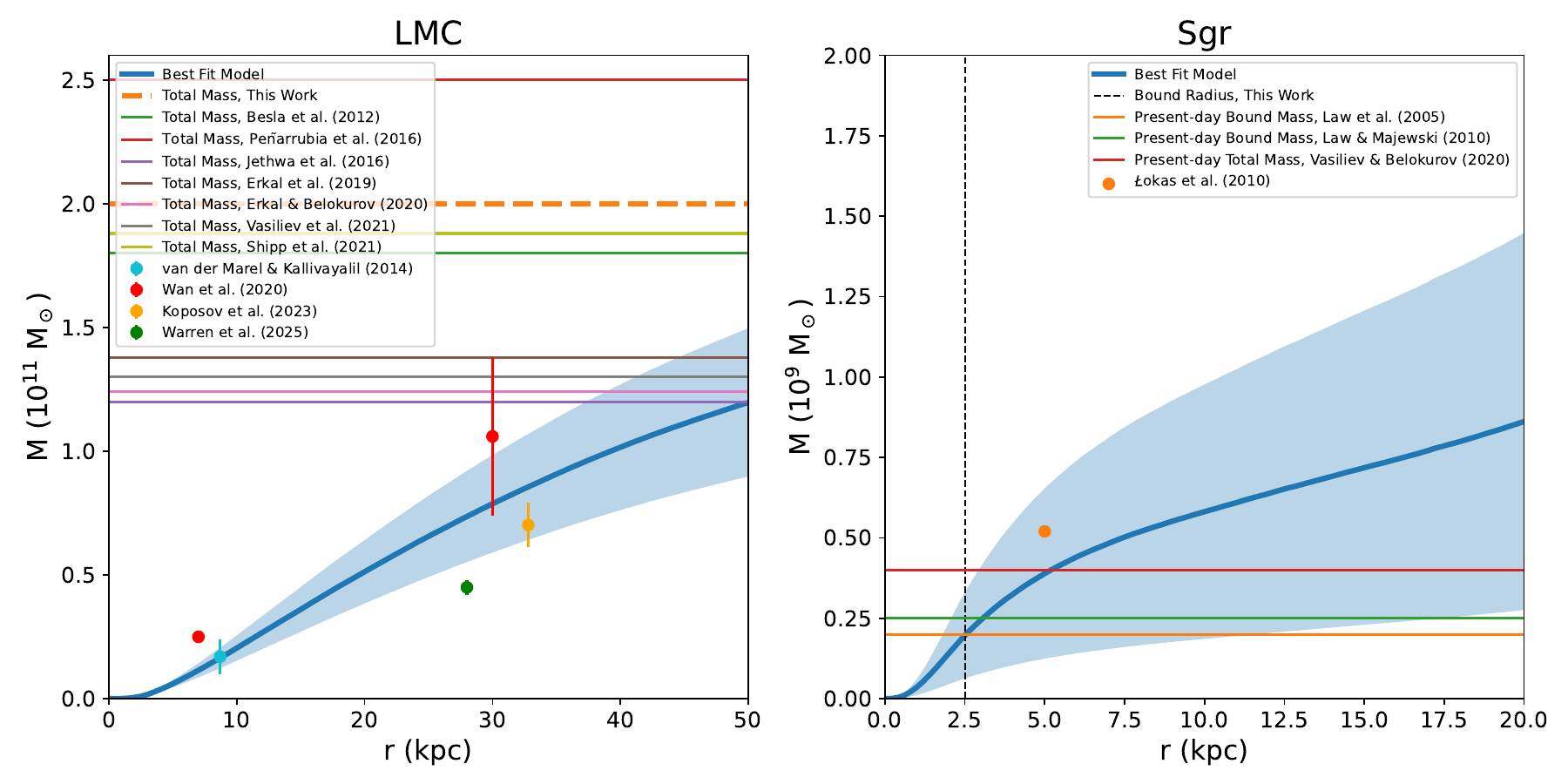}
    \caption{Mass profiles as a function of radius for the best-fit satellite models. Also shown are various literature values for the masses; bound and/or total masses are shown as horizontal lines, and masses enclosed within some radius are shown as points with error bars. Our determination of the LMC mass broadly agrees with previous measurements of the enclosed mass at various radii. Although our best-fit total mass is somewhat large compared to the literature values, note that the definition of ``total mass'' is generally not the same across studies, which can lead to systematic differences.  Our best-fit Sgr mass is on the lower end of the literature values when comparing the bound mass of the satellite (mass enclosed within the vertical dashed line). Estimates for the total mass of Sgr (both present-day and original) are not shown here because they are far off the vertical scale of the plot (see Table \ref{tab:sat_masses}). }
    \label{fig:supp_mass_fr}
\end{figure*}

%See also section 6.1 of Hasselquist et al. the big apogee paper

Previous works have constrained the total masses of the LMC from kinematic data of stars and gas; these values are provided in Figure \ref{fig:supp_mass_fr} along with the radial mass distribution for the satellites at the end of our best-fit simulation, as well as Table \ref{tab:sat_masses}. Early studies on this topic showed that the morphology of stars and gas in the Magellanic system could be explained if the LMC had a total mass of 1.8$\times$10$^{11}$ M$_\odot$ \citep{Besla2012}. Somewhat later, it was pointed out that cosmological arguments for the observed velocities of local group galaxies required a much larger total mass of 2.5$^{+0.9}_{-0.8}\times10^{11}$ M$_\odot$ for the LMC \citep{Penarrubia2016}; simultaneously, the motions of MW dwarf galaxies implied a smaller LMC mass of 1.2$\pm0.8\times10^{11}$ M$_\odot$ \citep{Jethwa2016}, also see \citep{ErkalBelokurov2020}. Recently, estimates for a total mass of the LMC of 1.3$-$1.4$\times10^{11}$ M$_\odot$ have been made based on the dwarf's interactions with stellar streams and the MW inner halo/disk \citep{Erkal2019,Vasiliev2021}. Our inferred value for the total mass of the LMC is on the heavier side of these literature values, although it lies within the range of previous measurements. The mass profile of the LMC within a few dozen kpc agrees very well with literature values, however. Estimates for the original total mass of Sgr are predominantly based on the observed velocity dispersion of stars \citep{LawMajewski2010,Lokas2010} and the chemo-dynamical properties of the dwarf remnant \citep{Gibbons2017,Mucciarelli2017}, and range nearly two orders of magnitude from 6.4$\times10^8$ M$_\odot$ to 6$\times10^{10}$ M$_\odot$. Our inferred mass for Sgr lies within this range, and is somewhat smaller than the estimates from recent chemo-dynamic analyses.

We report the total mass of each satellite at the beginning of the simulation; however, as satellites become stripped and otherwise evolve throughout the simulation, their (bound) mass distribution can change dramatically compared to their original total mass, so the true present-day mass will be smaller than this value \citep{Kravtsov2004}. Characterizing the mass of a satellite is not always straightforward because dwarf galaxies do not have well-defined boundaries, and as such it is difficult to compare the mass of a dwarf galaxy in a simulation to observations of real dwarf galaxies. In simulations, one has access to the true underlying mass distributions of a dwarf galaxy at all distances; in real life, we are limited in what material we can observe within dwarf galaxies, as well as what stripped material can be reasonably assigned as originating from a specific dwarf galaxy, and we rely on indirect measurements of the dark matter content of those objects. As a result, when discussing dynamical analyses, the inferred ``total mass'' of a satellite depends strongly on the chosen mass profile for that satellite -- regardless of data and procedure -- because the quantity that is actually being probed is the mass enclosed within some radius \citep{Warren2025}. Including simulated LMC particles out to over 100 kpc from the dwarf galaxy's apparent center as part of the ``total mass'' is clearly not a realistic representation that data when comparing to observationally derived quantities. On the other hand, estimates of dwarf galaxy masses based on the chemical abundances of their member stars are expected to reflect the true total mass of the original bound satellite \citep{Kirby2013}. Similarly, it is difficult to pick a specific way to compare the mass of a satellite to observations, since each observational method is sensitive to the mass within a different range of radii.

A more practical measurement of the satellite mass is the total (dark + baryon) mass enclosed within the tidal radius at the present day, which for the best-fit LMC is 4.1 $\pm$ 1.0 $\times$ 10$^{10}$ M$_\odot$ within a radius of 16.6 kpc, and for Sgr is 3.5 $\pm$ 2.4 $\times$ 10$^8$ M$_\odot$ within a radius of 5 kpc. Overall, it is clear that our method is primarily sensitive to the mass of the LMC, which is the main driver of the acceleration asymmetry near the Sun. However, we are also able to constrain the mass of the Sgr dwarf to within better than 1 sigma. For maximum clarity and flexibility, Figure \ref{fig:supp_mass_fr} shows the mass distributions as a function of radius for each satellite at the end of the best-fit simulation.

\subsection{On the Best-Fit Mass for Sagittarius} \label{sec:sgr_mass}

\begin{figure*}
    \centering
    \includegraphics[width=\linewidth]{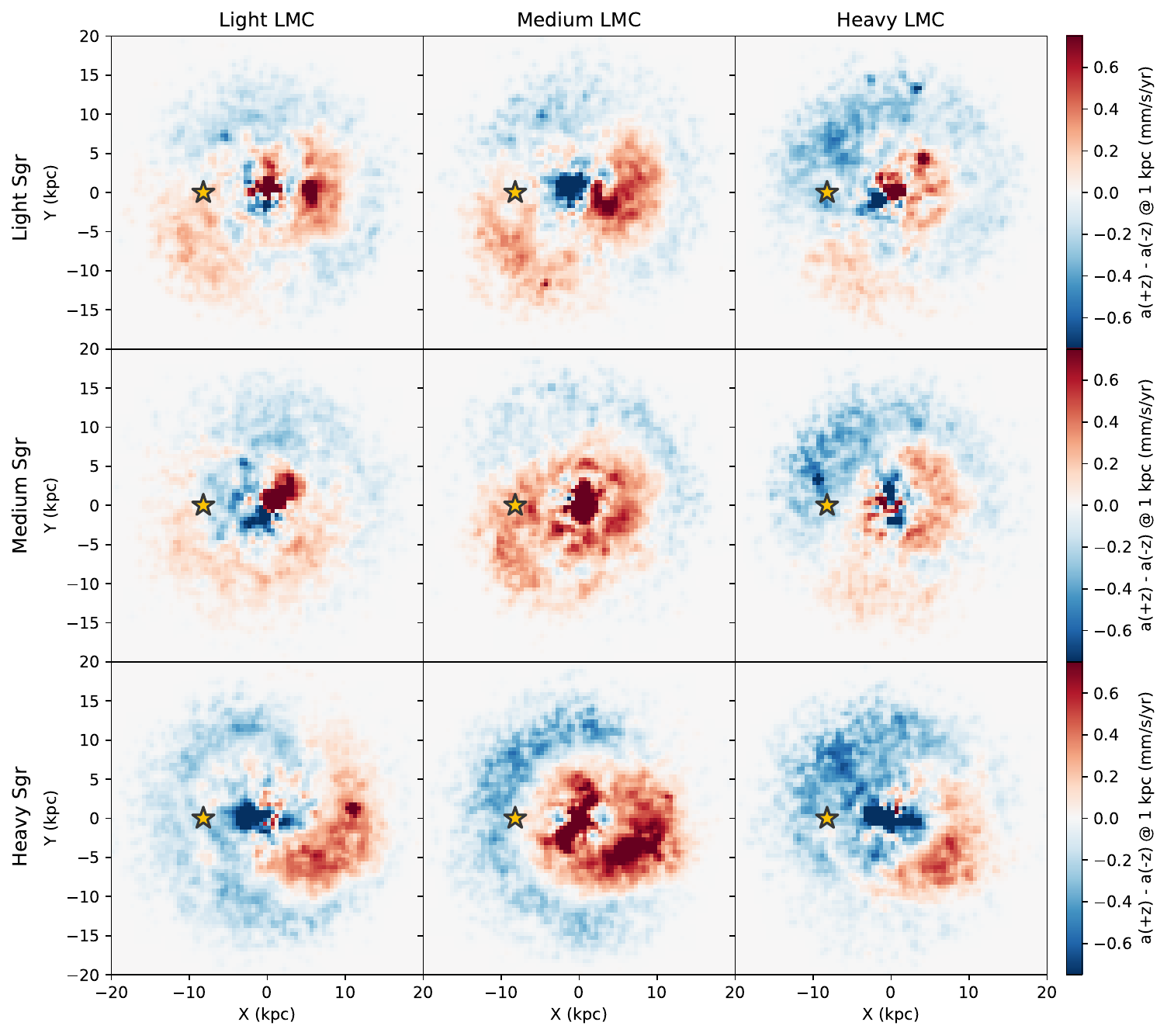}
    \caption{The vertical acceleration asymmetry across a face-on projection of the disk for simulations with different masses of the LMC and Sgr dwarfs. Columns and rows are the same as in Figure \ref{fig:supp_asym_grid}. Each panel shows the simulated acceleration asymmetry (acceleration above the disk minus acceleration below the disk) as color, where positive asymmetry is shown in red, and negative asymmetry (like that observed at the Sun) is blue. The location of the Sun is shown as a gold star. The structure induced by the two satellites is complicated and nonlinear, although there are potentially patterns in the asymmetries along the same rows or columns.}
    \label{fig:supp_asym_faceon}
\end{figure*}

\begin{figure*}
    \centering
    \includegraphics[width=\linewidth]{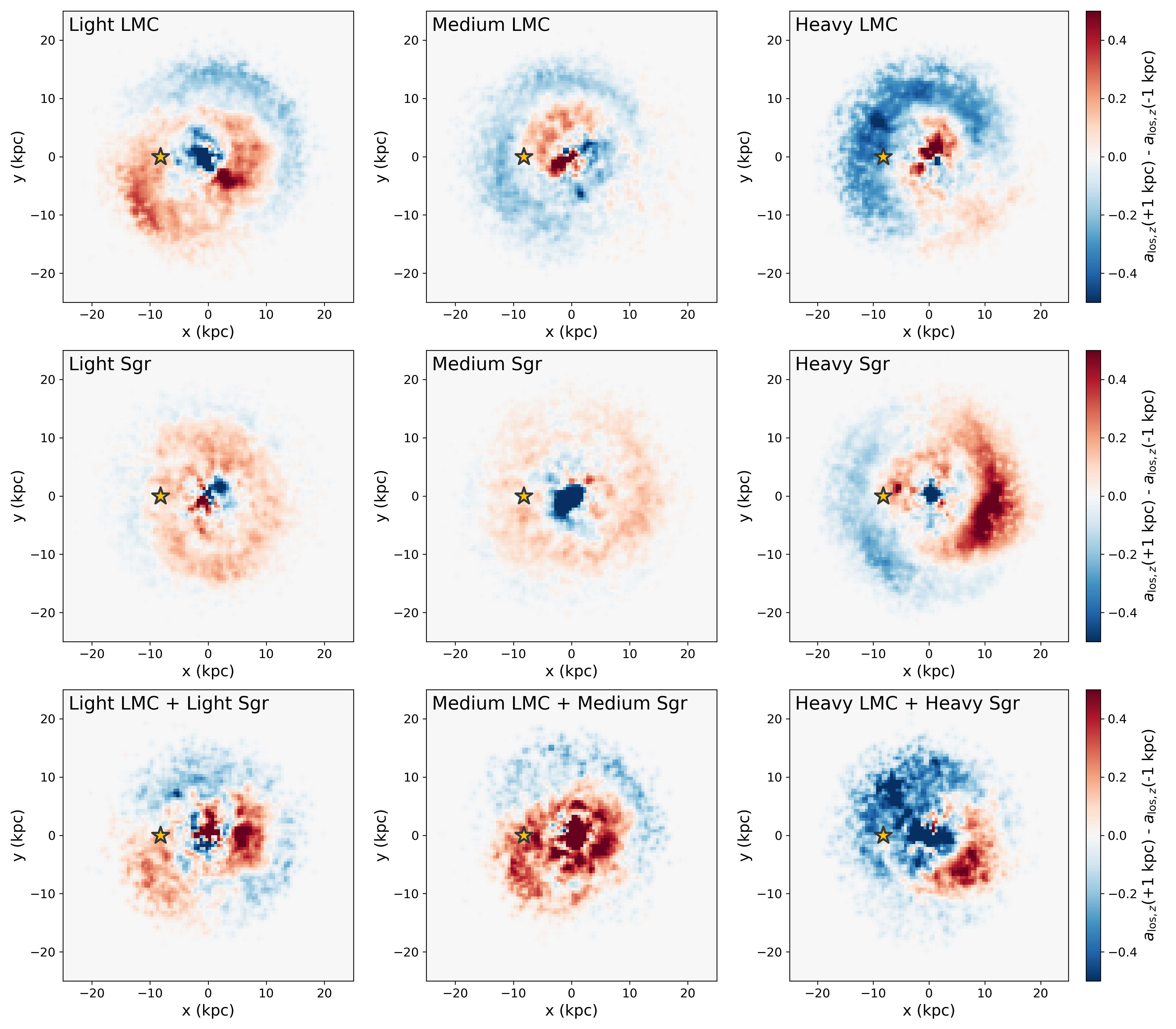}
    \caption{The vertical acceleration asymmetry across a face-on projection of the disk for simulations with different combinations of the LMC and Sgr dwarfs. The top row shows simulations with only the LMC, the middle row shows simulations with only Sgr, and the bottom row shows simulations with both dwarfs. The mass of each dwarf increases from left to right across columns. The location of the Sun is shown as a gold star. The effect of increasing a satellite's mass is not linear, in that a larger mass does not correspond to stronger accelerations in identical locations, but rather it changes both the relative locations and magnitudes of the acceleration asymmetries across the disk. Also note that the accelerations do not add linearly; the accelerations observed in a simulation with only the LMC plus the accelerations from a Sgr-only simulation do not necessarily correspond to the accelerations observed in a simulation with both of those dwarfs, even if they have the same mass. }
    \label{fig:lmc_sgr_comp}
\end{figure*}

We have previously pointed out that the asymmetries generated by orbiting satellites are sensitive to the mass of the satellite. Naively, one might expect a more massive satellite would always correspond to a larger asymmetry at the location of the Sun at the present day. Perhaps surprisingly, this is not the case. The right-most column of Figure \ref{fig:supp_asym_grid} shows that if one holds the mass of the LMC constant, increasing the mass of Sgr initially increases the magnitude of the perturbation; as the mass of Sgr is increased further, the observed magnitude of the asymmetry actually decreases.

The physical reasoning behind this is twofold. First, while a more massive satellite corresponds to a greater perturbation magnitude as a whole, it does not guarantee a greater perturbation magnitude at all positions on the disk simultaneously. This is because the perturbation is not uniform across the disk; it is composed of many individual substructures, such as waves, warps, etc. that each interact with each other and contribute to the overall time evolution of the perturbation. Second, a more massive satellite will experience more dynamical friction, which will change the travel time of the satellite over its orbit.  As the mass of the satellite -- and therefore its position at a given time -- is changed, the host galaxy is affected by torques at a different time. As there are multiple dwarf galaxies present in these simulations, the torques between the satellites may add together or partially cancel out depending on where the two satellites are relative to one another, which will change the mean acceleration felt by different parts of the disk at a given time. 

This matters little for the effect of the LMC, which only experiences a single pericentric passage in the simulation. However, the Sgr dSph orbits the galaxy three times throughout the course of the simulation. As the host galaxy model is the same in all simulations, the dynamical timescales of the host galaxy disk and halo remain fixed, but the time that the Sgr dSph reaches pericenter will be different. As a result, the disk rotates and phase mixes a different amount between pericentric passages for the different Sgr models, leading to a complicated overall effect on the present-day accelerations in the disk, as shown in Figure \ref{fig:supp_asym_faceon}. This complicated effect, combined with the fact that Sgr appears to have a somewhat smaller impact on disk accelerations than the LMC because of its smaller mass (Figure \ref{fig:lmc_sgr_comp}), leads to a larger relative uncertainty on the mass of Sgr using our method. 

% The simulation also generates a phase space spiral near the Sun in the best-fit simulation, which is shown in Figure \ref{fig:supp_pss}. The observed spiral is not as tightly wound as those observed from the Gaia data \citep{Antoja2018} due to the limited force resolution of the simulations. Additionally, it appears to be rotated roughly 90$^\circ$ counter-clockwise compared to the observed Gaia phase space spiral. 

\subsection{Accelerations Across the Disk}

\begin{figure}
    \centering
    \includegraphics[width=\linewidth]{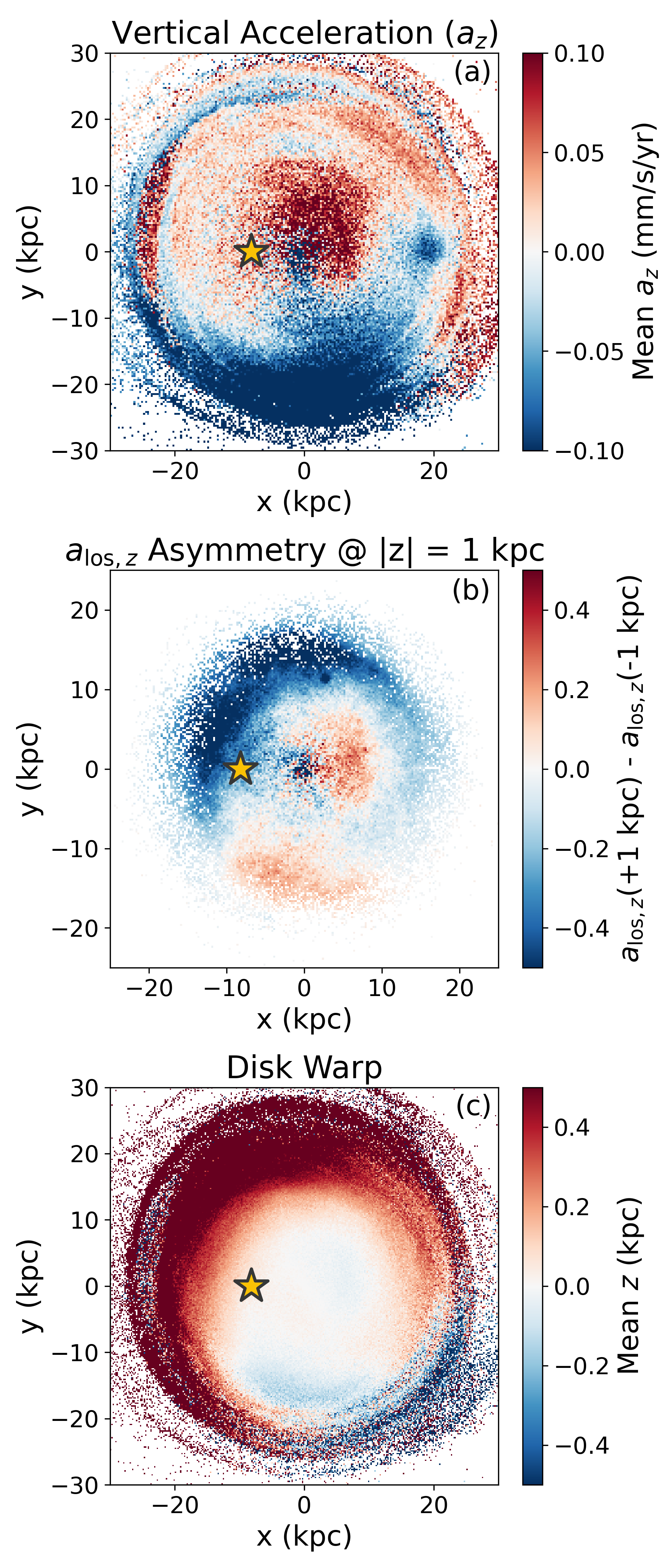}
    \caption{Vertical disk structure in the best-fit simulation. All panels show a face-on projection of the disk, and the location of the Sun is shown as a golden star. Panel (a) shows the mean vertical acceleration across the disk. Panel (b) shows the vertical acceleration asymmetry across the disk, calculated as the difference in $a_{\mathrm{los},z}$ at $z=1$ kpc and $z=-1$ kpc from the midplane. Panel (c) shows the mean height of the disk stars (i.e. the warp).}
    \label{fig:accels_and_warp}
\end{figure}

Up to this point, we have only considered the acceleration asymmetry near the Sun, where the pulsar data is located. However, because we simulate the entire Galaxy, we are able to analyze accelerations across the entire disk. Figure \ref{fig:accels_and_warp} shows the vertical acceleration, acceleration asymmetry, and mean vertical height of the disk as a function of position for the best-fit simulation. The vertical acceleration profile in panel (a) shows prominent ring-like structures, similar to the findings of \cite{Donghia2016}. Note that one side of the disk (towards the Sun's motion) is accelerating upwards, while the other half of the disk is accelerating downwards -- in other words, the disk is actively tilting. This effect is due to the passage of the LMC, which generates tidal forces across the MW disk \citep{Besla2007,Laporte2018}. The disk warp appears to roughly follow the same pattern as the acceleration field, although it is rotated somewhat in the direction opposite the disk rotation. This is because the warp is generated from the accelerations felt by the disk, but lags behind the true present-day acceleration because it takes time for an impulse to generate a significant shift in the velocities and positions of the disk stars. The location of the warp in the best-fit simulation roughly agrees with the observed location of the warp in the MW disk \citep{Skowron2019}, even though we did not optimize the simulations to fit the warp in any way. This showcases how the direct acceleration measurements contain a significant amount of dynamical data, especially when one considers the relatively small number of individual measurements and small region of the disk probed by the current direct acceleration data. 

It should be noted that there is a general consensus that the LMC alone cannot be responsible for the observed MW disk warp due to differences in the line-of-nodes and amplitudes of the warp in the outer disk \citep{Garcia-Ruiz2002,Laporte2018}. Although disk waves and the warp do contribute roughly 25\% of the asymmetry (the exact contribution depends on distance from the midplane), the majority of this effect comes from changes in the centers of mass between the inner dark matter halo and the baryonic disk. As a result, even if the warp in the simulation does not match the observed MW disk warp exactly, this effect is subdominant to the overall accelerations induced by the LMC in the dark matter halo, and does not seriously impact the quality of our fit or results.

The acceleration asymmetry in panel (b) of Figure \ref{fig:accels_and_warp} is roughly opposite to the mean height of disk stars at any point in the disk (an instantaneous local displacement towards negative $z$ will lead to an overall restoring force upwards towards the mean midplane). There are also strong accelerations in the inner few kpc of the disk due to the formation and spin of a bar, although these accelerations do not extend out to the Sun's orbital radius. While processes such as bar buckling can generate vertically asymmetric structure in the disk \citep{Lokas2019}, a stable bar like the one in these simulations will not produce any asymmetry in vertical accelerations and can be ignored for the purposes of this study.  %Note that the acceleration asymmetry is not due to shifts in the local midplane height of the disk, because we center the disk before calculating accelerations. 

We emphasize that this approach is able to produce information about dwarf galaxies which are tens of kpc away despite only using pulsar acceleration from within a few kpc of the Sun. This is possible because the effects of those satellite galaxies have significant impacts throughout the entire Galaxy, which are not localized only to regions near the dwarf galaxies. For example, tidal effects from the LMC result in the MW disk being displaced from the center of mass of the MW's dark halo; although this is a large-scale global effect, because the accelerations across the entire disk are impacted according to a specific pattern, one only needs access to accelerations from a dynamically small region of the disk to constrain the global feature. This is similarly true for Sgr, although the pattern of accelerations induced by each satellite are different due to the disparate orbits of the two dwarf galaxies. %The small dynamic range of the pulsar data can be compared to kinematic studies, which often rely on measurements of stars and gas that are located at sizable distances from the Sun in order to obtain mass estimates for these dwarf galaxies. However, the small dynamical range of the pulsar data is similar to the extent of data in studies of the vertical phase space spiral \citep{Antoja2018}, which can potentially constrain properties of the Sgr dwarf galaxy from only stars near the Sun \citep{Laporte2019}, although simulations of Sgr have so far been unable to reproduce all observed features of the phase space spiral \citep{Antoja2023}.

\subsection{Triaxial Halo}

\begin{figure}
    \centering
    \includegraphics[width=\linewidth]{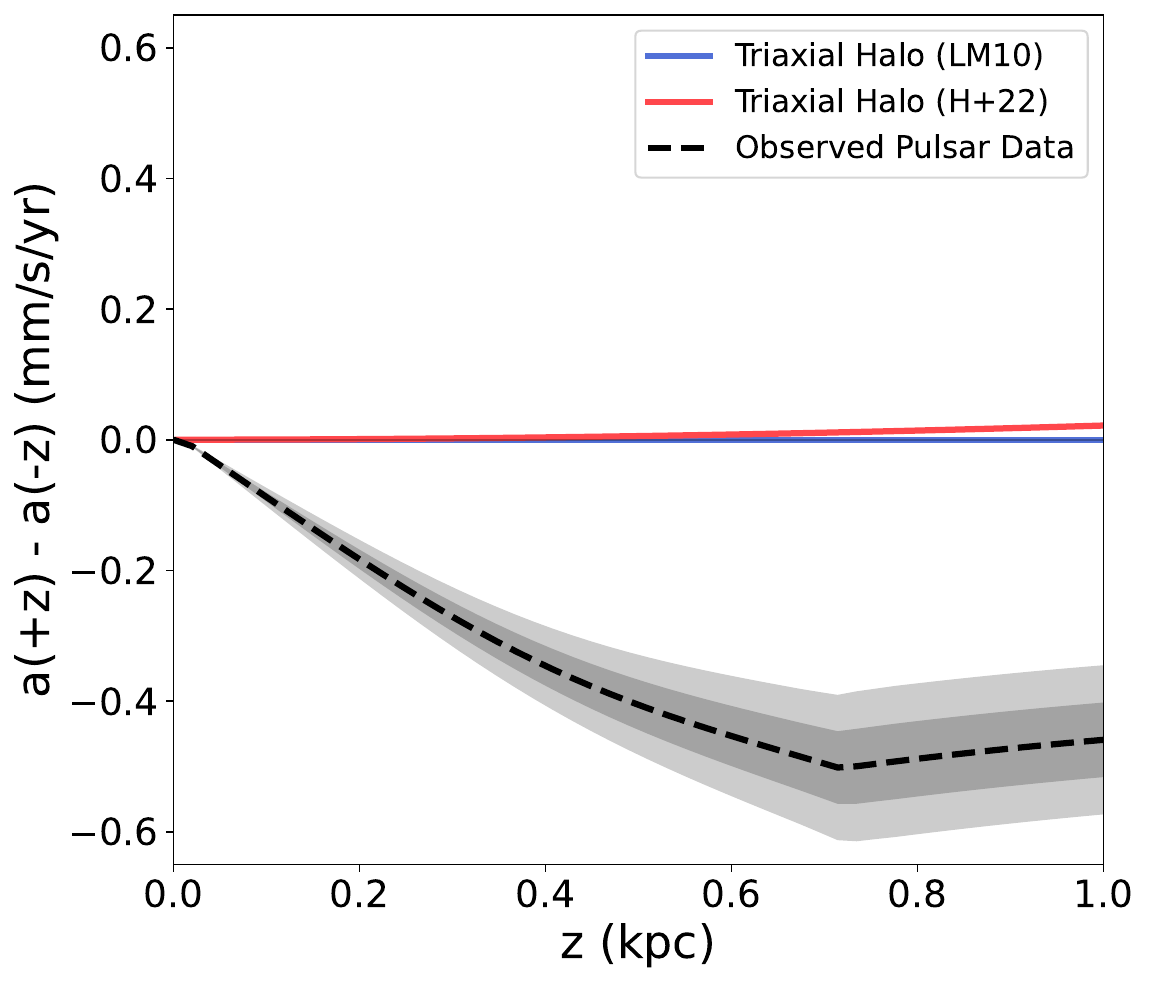}
    \caption{The observed vertical acceleration asymmetry (dashed black line, 1 and 2$\sigma$ uncertainty regions shown as the gray bands) and the prediction for the acceleration asymmetry produced by two different (tilted) triaxial halo models by \cite{LawMajewski2010} and \cite{Han2022a}. The magnitude of the asymmetry from the triaxial potentials is much smaller than the observed signal, and as a result is not expected to significantly contribute to accelerations near the Sun. }
    \label{fig:triaxial}
\end{figure}

Modern cosmological simulations have shown that the dark matter halos of galaxies are generally not spherical, but stretched in one or more directions. This can arise due to the infall of matter along filamentary lines \citep{Schneider2012} as well as changes in the halo morphology due to merger events \citep{Baptista2023}. Early dynamical evidence for the halo being non-spherical included simulations of the Sgr stream by \cite{LawMajewski2010}, who found that a tilted and triaxial dark matter halo was required to match the observed tidal features of the stripped dwarf galaxy. Recently, a series of papers \citep{Han2022a,Han2022b,Han2023,Han2023Nat} have shown that the distribution of stars in the MW halo follow a tilted triaxial distribution; because the tilts and shapes of dark matter halos tend to match the properties of the embedded stellar halo in simulations, this is strong observational evidence that the dark matter halo of the MW is in-fact tilted and shaped like a spheroid, rather than a sphere. 

As a tilted triaxial dark matter halo profile places a different amount of mass on either side of the disk, these models will lead to an acceleration asymmetry. The asymmetry predicted by two independent triaxial halo models, \cite{LawMajewski2010} and \cite{Han2022a}, has the opposite sign as the asymmetry obtained using direct acceleration measurements of pulsars, and a much smaller magnitude (Figure \ref{fig:triaxial}). As a result, we do not expect the large-scale triaxiality of the dark matter halo to significantly contribute to the observed accelerations near the Sun, and is unlikely to affect the results of this work.

% This discrepancy could be because the triaxial halo models are fit to data further out from the Sun -- \cite{LawMajewski2010} used the morphology of the Sgr stream at a few tens of kpc, and \cite{Han2022a} likewise analyzed the density of stars across the entire halo -- whereas our pulsar data only spans a few kpc from the Sun. It could very well be the case that there is more mass below the Galactic midplane close to the Sun, as predicted from the pulsar data, and that at larger distances there is more mass above the midplane. Additionally, \cite{LawMajewski2010} suggested that the modeled triaxiality of the halo may disappear when one accounts for the additional mass brought in with the LMC. This inconsistency between studies of the tilted triaxial halo and studies of the pulsar data provides an interesting avenue for future work. 

\section{Conclusion}

We have outlined here a method for measuring the masses of the LMC and Sgr dwarf galaxies in a way that is entirely independent of any kinematic stellar data; this new approach relies only on observations of gravitational accelerations. These orbiting satellites generate disequilibrium effects, which propagate through the entire Galaxy; the magnitude and relative orientation of these are related to the mass of each dwarf galaxy. 

By using direct acceleration data obtained from pulsars, we identify a vertical acceleration asymmetry around the Sun, which is the local manifestation of these global disequilibrium effects. Observing this disequilibrium requires precise temporal resolution that are not achievable with kinematic data (as the accelerations of interest disappear within one dynamical time), making direct acceleration measurements particularly suitable for this task.

We run a suite of simulations of a MW-like host galaxy plus dwarf galaxies that correspond to the LMC and Sgr. In each simulation, the final location and velocity of each dwarf galaxy corresponds to the observed present-day information for the corresponding satellite. We vary the masses of the two satellites in these simulations, and then determine the vertical acceleration asymmetry produced in each simulation at the location of the Sun. By comparing the simulations to the observed pulsar acceleration data, we are able to constrain the masses of the two satellite galaxies simultaneously. We determine the total mass of each satellite 3 Gyr ago to be 2.0 $\pm$ 0.5 $\times$ 10$^{11}$ M$_\odot$ for the LMC, and 4.4 $\pm$ 3.1 $\times$ 10$^{9}$ M$_\odot$ for the Sgr dSph. At the present day, the mass enclosed within the tidal radius of each satellite is 4.1 $\pm$ 1.0 $\times$ 10$^{10}$ M$_\odot$ within a radius of 16.6 kpc for the LMC, and 3.5 $\pm$ 2.4 $\times$ 10$^8$ M$_\odot$ within a radius of 5 kpc for Sgr. The uncertainty in the LMC's total mass is competitive with the precision obtained using kinematic methods, although our estimate for the mass of Sgr is less precise. 

As this is the first application of pulsar accelerations to satellite mass inference, we have demonstrated that the direct acceleration data encodes enough information to constrain the properties of distant satellite galaxies. There are still several avenues for improving this method, which are likely to improve the precision of the satellite mass estimates, and could also feasibly provide information about additional parameters such as the shape of these dwarf galaxies or the structure of the MW. Future studies which directly fit the 3-dimensional pulsar data, explore the parameter space in greater detail, and/or implement a fully Bayesian analysis may be able to significantly improve upon the constraints we provide here, or perhaps provide constraints for even more satellite galaxies. 

We remind the reader that the procedures used in this study are a fairly idealized treatment of a particularly complex problem. Essentially, one is tracing structures backwards in time through a dissipative system in order to predict their effects at the present day. The exact orbits, shapes, and mass-loss histories of these structures depend on the choice of initial conditions plus complicated dynamical effects which are known to be present in the MW but were not (directly) considered in this work. These include, but are not limited to: orbital resonances between halo dark matter structures and the satellite orbits; interactions with other satellites; resonances with the bar; different models for the MW mass distribution; different dark matter models; etc. As the satellites interact with each other gravitationally, any changes in these properties feed back into changes into the other (dwarf) galaxies, leading to a complicated non-linear system. These processes could all meaningfully change our findings, and should be thoroughly explored in the future in order to obtain definitive constraints on the masses of these dwarf galaxies.

With that said, this proof of concept illustrates the power of only a few direct acceleration measurements over a relatively small dynamic range to constrain large-scale features of our Galaxy and its satellites, which otherwise require a substantial number of kinematic and/or chemical measurements of stars and gas. This result pushes us towards a framework of Galactic structure and dynamical modeling that jointly incorporates the real-time accelerations of objects as well as the kinematics of stars \citep{Craig2023}. More accelerations are expected to rapidly become available, both from pulsar timing as well as other precision measurements including extreme-precision radial velocity observations \cite{Chakrabarti2020} and eclipse timing observations \cite{Chakrabarti2022}. Additionally, existing pulsar acceleration measurements quickly become more precise, favorably scaling as a function of the observational baseline time \citep{Chakrabarti2021,Donlon2024}. For these reasons, it is important that we work to combine these novel interdisciplinary tools with the historical tools of Galactic science, as it is likely to enable exciting new discoveries that would not be possible otherwise.

\acknowledgments

We would like to thank Lars Hernquist for helpful conversations. SC acknowledges support from NASA EPSCoR CAN AL-80NSSC24M0104, STSCI GO 17505, and the Margaret Burbidge fellowship.
JH is supported by a UKRI Ernest Rutherford Fellowship ST/Z510245/1.
Computational support was provided by the RCHAU project, NSF grant 2232873.

\bibliographystyle{aasjournal}
\bibliography{references.bib}

\appendix

\section{Initial Conditions of Simulations}

\begin{table*}[]
    \centering
    \footnotesize
    \begin{tabular}{lccccr} \hline
         Simulation & LMC $\mathbf{x}$ & LMC $\mathbf{v}$ & Sgr $\mathbf{x}$ & Sgr $\mathbf{v}$ & Runtime \\
         & (kpc, kpc, kpc) & (km/s, km/s, km/s) & (kpc, kpc, kpc) & (km/s, km/s, km/s) & (Gyr) \\ \hline \hline
         Light LMC \& Light Sgr & (25.31, 339.02, 97.83) & (5.58, -43.14, -61.20) & (66.34, 2.36, -26.21) & (9.84, -45.98, 91.89) & 3.0 \\ 
         Light LMC \& Medium Sgr & (25.31, 339.02, 97.83) & (5.58, -43.14, -61.20) & (66.34, 2.36, -26.21) & (9.84, -45.98, 91.89) & 3.0 \\
         Light LMC \& Heavy Sgr & (26.92, 324.51, 78.72) & (4.90, -51.66, -63.45) & (66.34, 2.36, -26.21) & (9.84, -45.98, 91.89) & 2.7 \\
         Medium LMC \& Light Sgr & (24.73, 343.29, 104.05) & (5.79, -40.55, -60.39) & (66.34, 2.36, -26.21) & (9.84, -45.98, 91.89) & 3.0 \\ 
         Medium LMC \& Medium Sgr & (25.31, 339.02, 97.83) & (5.58, -43.14, -61.20) & (66.34, 2.36, -26.21) & (9.84, -45.98, 91.89) & 2.9 \\
         Medium LMC \& Heavy Sgr & (26.41, 329.64, 85.17) & (5.14, -48.71, -62.71) & (66.34, 2.36, -26.21) & (9.84, -45.98, 91.89) & 2.7 \\
         Heavy LMC \& Light Sgr & (24.13, 347.36, 110.18) & (5.97, -38.01, -59.60) & (66.34, 0.00, -21.33) & (-10.12, -46.28, 98.95) & 3.0 \\ 
         Heavy LMC \& Medium Sgr & (24.73, 343.29, 104.05) & (5.79, -40.55, -60.39) & (66.34, 2.36, -26.21) & (9.84, -45.98, 91.89) & 2.9 \\
         Heavy LMC \& Heavy Sgr & (26.14, 332.09, 88.36) & (5.25, -47.28, -62.34) & (65.35, 4.68, -30.71) & (28.69, -44.88, 83.81) & 2.7 \\
         M.-Heavy LMC \& M.-Heavy Sgr & (25.31, 339.02, 97.83) & (5.58, -43.14, -61.21) & (66.34, 0.00, -21.33) & (-10.12, -46.28, 98.95) & 2.8 \\
         M.-Heavy LMC \& Medium Sgr & (24.73, 343.29, 104.05) & (5.79, -40.55, -60.39) & (66.34, 0.00, -21.33) & (-10.12, -46.28, 98.95) & 2.9 \\
         M.-Heavy LMC \& M.-Heavy Sgr & (24.73, 343.29, 104.05) & (5.79, -40.55, -60.39) & (63.42, 6.93, -34.76) & (46.77, -43.17, 74.74) & 2.9 \\
         Heavy LMC \& M.-Heavy Sgr & (24.73, 343.29, 104.05) & (5.79, -40.55, -60.39) & (66.34, 2.36, -26.21) & (9.84, -45.98, 91.89) & 2.9 \\
         Heavy LMC \& M.-Heavy Sgr & (24.73, 343.29, 104.05) & (5.79, -40.55, -60.39) & (63.42, 6.93, -34.76) & (46.77, -43.17, 74.74) & 2.9 \\
         H.-Plus LMC \& M.-Heavy Sgr & (24.44, 345.33, 107.12) & (5.88, -39.27, -60.00) & (65.35, 4.68, -30.71) & (28.69, -44.88, 83.81) & 2.95 \\ 
         H.-Plus LMC \& Medium Sgr & (24.73, 343.29, 104.05) & (5.79, -40.55, -60.39) & (66.34, 2.36, -26.21) & (9.84, -45.98, 91.89) & 2.9 \\
         H.-Plus LMC \& M.-Heavy Sgr & (24.73, 343.29, 104.05) & (5.79, -40.55, -60.39) & (63.42, 6.93, -34.76) & (46.77, -43.17, 74.74) & 2.9 \\\hline
    \end{tabular}
    \caption{Initial positions and velocities of the LMC and Sgr in the various simulations from this work, plus the runtime of each simulation.}
    \label{tab:ics}
\end{table*}

In Table \ref{tab:ics}, we provide the initial positions and velocities of the satellites relative to the MW in each simulation. The masses and shapes of the various simulation components are given in Table \ref{tab:model_params} and Section \ref{sec:simulations}. 

\section{Previous Measurements of Satellite Masses}

\begin{table*}[hbt!]
    \centering
    \begin{tabular}{lrr} \hline
        \textbf{LMC} & & \\ \hline
        Constraint or Radius (kpc) & Mass (10$^{11}$ M$_\odot$) & Reference \\ \hline \hline
        Total Mass & 1.8 & \cite{Besla2012} \\
        8.7 & 0.17(7) & \cite{vanderMarelKallivayalil2014} \\
        Total Mass & 2.5(9) & \cite{Penarrubia2016} \\
        Total Mass & 1.2(8) & \cite{Jethwa2016} \\
        Total Mass & 1.38(27) & \cite{Erkal2019} \\
        7 & 0.25(1) & \cite{Wan2020} \\
        30 & 1.06(32) & \cite{Wan2020} \\
        Total Mass & $>$1.24 & \cite{ErkalBelokurov2020} \\
        Total Mass & 1.3(3) & \cite{Vasiliev2021} \\
        Total Mass & 1.88(40) & \cite{Shipp2021} \\
        32.8 & 0.702(90) & \cite{Koposov2023} \\ 
        28 & 0.45(3) & \cite{Warren2025} \\ \hline
        
        \textbf{Sgr} & & \\ \hline
        Constraint or Radius (kpc) & Mass (10$^{8}$ M$_\odot$) & Reference \\ \hline \hline
        Present-day Bound Mass & 2$-$5 & \cite{Law2005} \\ 
        Present-day Bound Mass & 2.5(13) & \cite{LawMajewski2010} \\
        Original Bound Mass & 6.4 & \cite{LawMajewski2010} \\
        5 & 5.2 & \cite{Lokas2010} \\
        Original Total Mass & 160 & \cite{Lokas2010} \\
        Original Total Mass & $>$600 & \cite{Gibbons2017} \\
        Original Total Mass & 600 & \cite{Mucciarelli2017} \\
        Present-day Total Mass & 300 & \cite{Bland-Hawthorn2019} \\
        Present-day Total Mass & 4 & \cite{VasilievBelokurov2020} \\ \hline
        
    \end{tabular}
    \caption{Various literature values for the masses of the LMC and Sgr dwarf galaxies.}
    \label{tab:sat_masses}
\end{table*}

We include here in Table \ref{tab:sat_masses} the data and references for previous constraints on the masses of the LMC and Sgr dSph. These values are also plotted in Figure \ref{fig:supp_mass_fr}.

\end{document}